# Ampere-class Bright Field Emission Cathode Operated at 100 MV/m

Mitchell Schneider,[1,2,3] Emily Jevarjian,[1,2] Tanvi Nikhar,[1] Taha Y. Posos,[1] Wanming Liu,[4] Jiahang Shao[4] and Sergey V. Baryshev[1]

1) Department of Electrical and Computer Engineering, Michigan State University, East Lansing, MI 48824, USA

2) Department of Physics and Astronomy, Michigan State University, East Lansing, MI 48824, USA

3) Accelerator Operations and Technology Division, Los Alamos National Laboratory, Los Alamos, NM 87545, USA

4) High Energy Physics Division, Argonne National Laboratory, Lemont, IL 60439, USA

**Abstract**

High current bright sources are needed to power the next generation of compact rf and microwave systems. A major requirement is that such a source could be sustainably operated at high frequencies, well above 1 GHz, and high gradients, well above 100 MV/m. Field emission sources offer simplicity and scalability in a high frequency era of the injector design, but the output rf cycle charge and high gradient operation remain a great and largely unaddressed challenge. Here, a field emission cathode based on ultra-nano-crystalline diamond or UNCD, an efficient planar field emission material, was tested at 100 MV/m in an L-band injector. A very high charge of 38 pC per rf cycle (300 nC per rf pulse corresponding to rf pulse current of 120 mA) was demonstrated. This operating condition revealed a space charge dominated emission from the cathode and revealed a condition under which the 1D Child Langmuir limit was surpassed. Finally, a beam brightness of ~$10^{14}$–$10^{15}$ A/m²×rad² was estimated.

**Introduction**

Rf injectors are the working horse electron guns producing relativistic electron beams finding ubiquitous applications in industrial, security, environmental, medical and basic science sectors. The performance comparison can be drawn in terms of output current $I$, current density $j$, or beam brightness $B$ which are the most challenging and ambitious metrics to achieve. Brightness is defined as $B = \frac{2I}{\varepsilon_\perp^2}$, where the total normalized transverse emittance $\varepsilon_\perp$ is found as $\varepsilon_\perp^2 = \varepsilon_{int}^2 + \varepsilon_{sc}^2 + \varepsilon_{rf}^2$ with $\varepsilon_{int}$ being the intrinsic cathode emittance, and $\varepsilon_{sc}$ and $\varepsilon_{rf}$ being the space charge induced and rf induced emittance, respectively. The definitions of the brightness and emittance set the stage for rf injector development. It thus involves material science, emission physics and high power rf design.

With no special means, such as cathode surface topography [1], the space charge term is reduced through the increase of the macroscopic cathode rf field $E$ (also termed gradient) as $\varepsilon_{sc} = \frac{1}{8} \frac{I}{I_A} \frac{\lambda}{\alpha} \frac{1}{3\frac{\sigma_\perp}{\sigma_z}+5}$, where $I_A$ is the Alfven current of 17 kA, $\lambda$ is the operating rf wavelength, $\sigma_\perp$ and $\sigma_z$ are transverse and longitudinal bunch sizes, respectively, and $\alpha = \frac{eE}{4\pi m_e c^2} \lambda$ with $m_e$ being the electron mass and $c$ being the speed of light [2]. Development of high frequency (X- to W- band) injectors is pivotal to greatly enhance the cathode gradient to above 100 MV/m and with 500 MV/m benchmark demonstrated, as higher operating frequency greatly suppresses the breakdown rate, greatly enhancing compactness at the same time.

One trade-off is that rf emittance grows with the gradient as $\varepsilon_{rf} = \frac{\sqrt{2}\pi^2}{\lambda^2} \frac{eE}{m_e c^2} \sigma_\perp^2 \sigma_z^2$ [3]. In the case of a photocathode, the rf emittance can be minimized by minimizing $\sigma_z$ through phase matched femtosecond laser. In the case of X-band frequency and well above [4], the use of photocathode is extremely challenging due to size constraints, and another trade-off of using a field emission cathode (FEC) in place of photo emission technology must be evaluated in great detail. To enable FEC operation, special dc-ac or harmonic mixing gating techniques or multi-cell designs were applied to reduce $\sigma_z$ [3,5,6]. Since an injector that features high *E* and reduced $\sigma_z$ has become available, it is critical to find a cathode material that 1) features low intrinsic emittance, 2) is capable of emitting 1-100 pC per rf cycle (translating to a current of many Amperes) and yet 3) is capable of surviving when exposed to gradient on the order of 100 MV/m and above.

The present work extends our prior developments toward FEC based high frequency rf injector technology utilizing ultra-nano-crystalline diamond (UNCD) as the cathode material. UNCD possess exceptional emission efficiencies [7-10] and low intrinsic emittance [11]. Diamond is a desired material for high power applications due to its thermal and mechanical properties. Despite this, diamond cathodes made in traditional high aspect ratio geometries, such as pyramids [12,13], cannot offer high gradient operation because they tend to explode at fields above 30-40 MV/m. Emission in diamond originates from sp2 grain boundaries [14,15], and UNCD has the large fraction of sp2 phase as sp3 grain size is small among polycrystalline diamond – these are key factors that allow for simple planar FEC geometry with ~10 nm roughness and promise operation at high gradients near or above 100 MV/m. Therefore, this study simultaneously assesses 1) survivability of UNCD to 100 MV/m and its overall conditioning dynamics, 2) ultimate output charge per rf cycle under ultimate gradient conditions, and 3) vacuum space charge related effects. The paper is laid out as the following: **Section II** describes the cathode fabrication; **Section III** describes the cathode testing facility; **Section IV** summarizes beam dynamics simulations and image processing approach; **Section V** summarizes conditioning procedure and cathode performance; **Section VI** provides discussion and physics implications; **Section VII** provides concluding remarks and outlook.

**Section II: Cathode Fabrication**

For this experiment, the UNCD cathode was grown on ultrasonically seeded molybdenum puck using microwave plasma assisted chemical vapor deposition in a S-band reactor operated at 2.45 GHz [16,17]. A synthesis substrate temperature of 1248 K was achieved using an $H_2/(20\%)N_2/(5\%)CH_4$ feed gas mixture maintained at a flow rate of 200 standard cubic centimeters per min (sccm) at a total gas pressure of 67.5 torr and 3000 W microwave power. The substrate temperature was measured using an infrared pyrometer during the 1-hour growth process. The UNCD coated puck was then mechanically attached to the three-part assembly, as outlined in Ref.[8]. Additionally, the edge of the cathode was carefully rounded and UNCD coated inner 18 mm of the 20 mm diameter puck to avoid edge effects.

The grown sample was then characterized using a Horiba Raman spectrometer with a 532 nm probing laser. The Raman spectrum typical to UNCD films illustrating D and G peaks centered around 1333 cm-1 and 1560 cm-1, respectively, were observed. This confirmed that the D peak corresponded to high fraction of sp3 diamond phase and G peak corresponded to semi-amorphous sp2 graphitic phases [16]. The deposition temperature was balanced such that the film exhibited high conductivity while still maintaining the sp3 phase (shown by the presence of 1333 cm-1 peak). From experiments on insulating Si and quartz witness substrates such a Raman corresponds to a resistivity of 0.5 Ω cm. As shown by the red line depicting the Raman taken before the experiment (Fig. 1), high conductivity is likely thanks to improved crystallinity and physical connectivity between sp2 grain boundaries as manifested by the G peak positioned at 1560 cm-1. Scanning Electron microscopy further confirms nanostructure typical for UNCD.

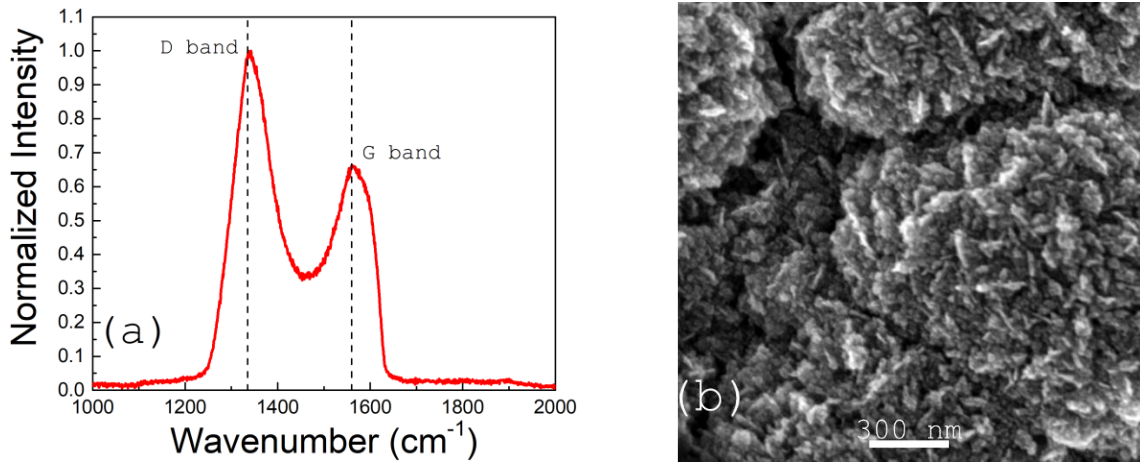

Figure 1: Raman spectrum and SEM micrograph of as-grown sample.

**Section III: Cathode Testing Facility**

An L-band (1.3 GHz) facility Argonne Cathode Test-stand (ACT), shown in Fig. 2, a part of the Argonne Wakefield Accelerator (AWA) switchyard was utilized for the fabricated UNCD FEC testing. This operating frequency allows for larger dimension cathodes, higher charges, and better imaging capabilities. From dc experiments [7,18], it is known that emission is non-uniform and non-monotonic with the macroscopic electric field. Therefore, ACT allows for extracting the charge and imaging the transverse emission profile to evaluate FEC performance and understand the underlying emission physics in the best possible ways.

The ACT injector is a single-cell normal conducting photoinjector [19] that is capable of operating in pure field emission regime with the laser turned off. The repetition rate of the system is 2 Hz, and the rf pulse length can be changed between 4 and 8 μs. The base vacuum pressure was maintained below $5 \times 10^{-9}$ Torr. The ACT's rf system also has a bidirectional coupler which is used to measure the forward and reverse power, each of which are attached to an attenuator located outside of the gun (not shown in Fig. 2).

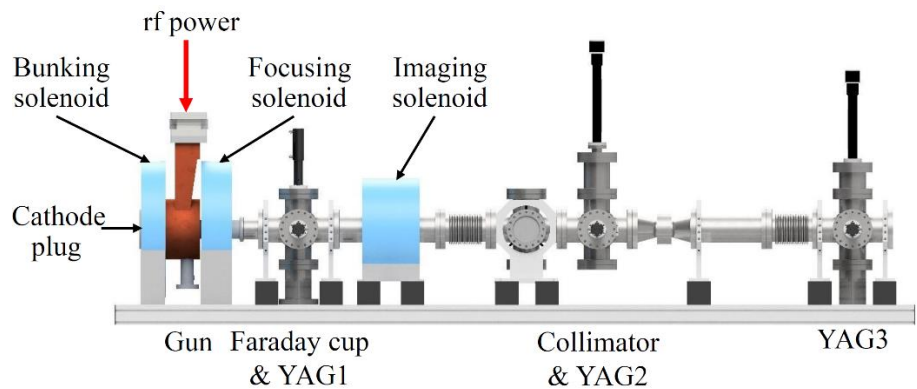

Figure 2: Schematic of the ACT beamline.

The ACT has three solenoids along the beamline. The first two are the bunking and focusing solenoids which are coupled together such that the current runs in opposite directions to cancel out the axial magnetic fields inside of the gun. Additionally, the focusing solenoid has more windings than the bunking solenoid, allowing for focusing of the beam immediately after it exits the gun. This focusing primarily serves the purpose of increasing the capture ratio, and previous results seen in Ref.[8] have shown that this ratio can be optimized to over 90% throughout conditioning.

*In situ* imaging is a unique feature of the ACT beamline: the third solenoid is the imaging solenoid, which is used to focus the beam for downstream imaging. To image the transverse electron distribution pattern, three Yttrium Aluminum Garnet (YAG) screens are used at different locations along the beam lines. YAG1 is used to image the emission pattern as the beam exits the gun and can be interchanged with the Faraday cup. YAG3, located 2.54 m away from the cathode surface, is used to image the downstream on-axis electrons by the addition of a 1 mm aperture at the location of YAG2. Images on YAG2 allow for optimization of the solenoid settings to produce a beam waste at the location of YAG2, 1.55 m from the cathode.

Combining the imaging and image processing (using FEpic [20]), beam dynamics (using FEgen [21] paired with GPT [22]), and the charge and field measurements that are derived from the Faraday cup and bidirectional coupler (all done through a pipeline FEbeam [23]) provides comprehensive analysis of the field emission cathode performance.

**Section IV: Image Processing and Beam Dynamics**

Imaging is supported by beam dynamics simulations as illustrated in Fig.3. The initial field emission distribution, 7×7 grid serving for better visualization, was created using a home-developed open-source software called FEgen which is explained in detail in Ref.[21] and then exported to GPT in order to further track the particles in three dimensions. The initial energy spread on the cathode surface was set to the intrinsic value of 0.1 eV, and the UNCD work function was assumed to be 4 eV from previous Kelvin probe measurements [24]. Fig.3 represents the experimental and simulation results at YAG2 and YAG3 positions, where the charge, local field, and solenoid settings were set to match the experimental settings corresponding to 70 MV/m. As the side-by-side comparison shows, the experimental results match the simulation results for a distributed emission from the cathode surface.

As seen from comparing simulations and imaging, emitters appear as bright stretched ellipses, but not circular spots like in a dc case [7,18]. The emission spots are stretched along rays, which start at the center of a cathode and go in all directions. They vary in length and brightness and are non-uniformly distributed in polar coordinates. Unlike in dc, electrons are generated by and interact with the rf/microwave drive cycle in a wide phase window. The extended interaction phase window causes the ellipses (or streaks) to form, are essentially represented as rotated projections of longitudinally stretched electron beamlets arriving from the cathode surface. Thus, it can be claimed that each line represents a singular emitter, and that counting streaks should be representative of the number of emitters and their variation as a function of the external power in the rf injector.

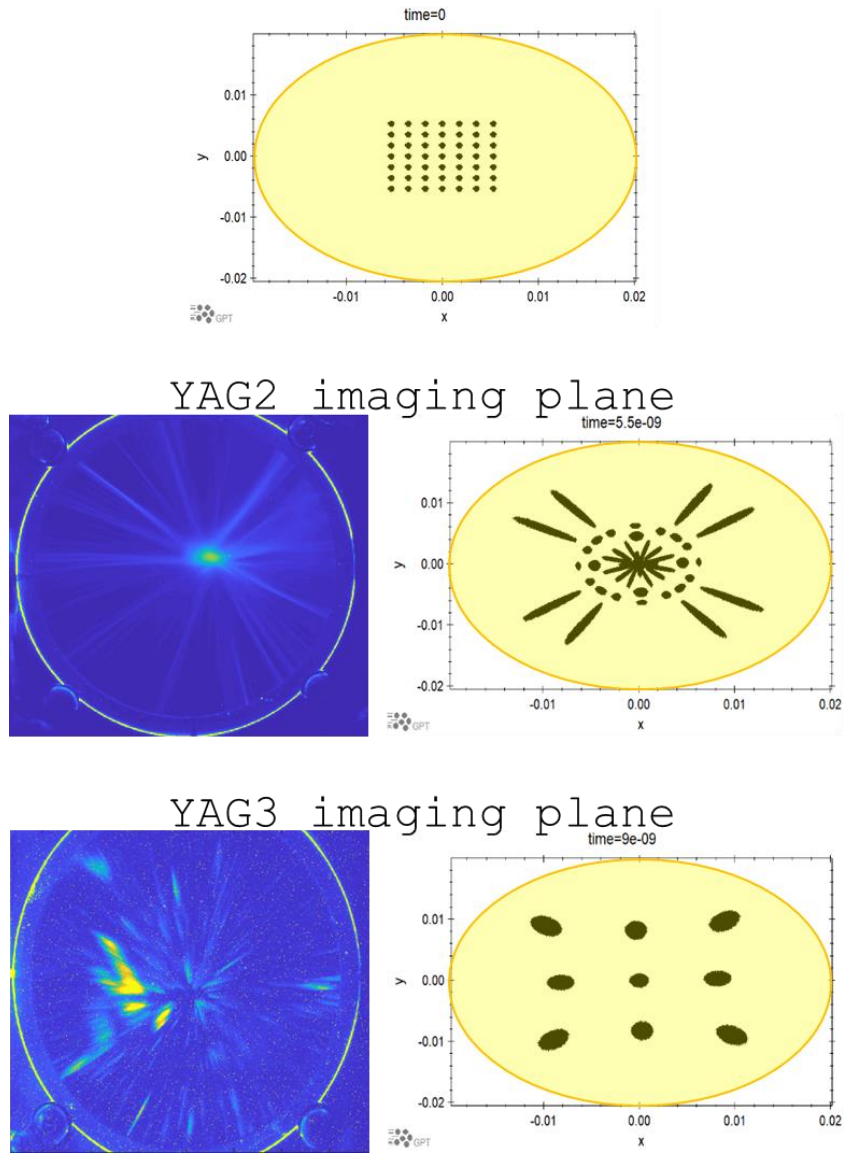

Figure 3: Comparison of transverse electrons patterns from experimental and simulated results extracted from GPT: initial emitter distribution 7×7 point like grid was launched in GPT and snap shots were taken at fly time stamps corresponding to YAG2 and YAG3 positions (left is actual emission micrographs and right is GPT results). The grid pattern seen in the simulation results match well with experimental measurements. Having a preserved grid pattern on YAG3 in the simulation results implies that, experimentally, there is a distributed emission from isolated emitters across the FEC surface. Note that each yellow circle represents the physical boundary of the YAG screens. *x* and *y* coordinates are in meters.

Charge collection was done hand in hand with image collection at different gradients. These collected images were processed using FEpic to obtain the emitter number statistics throughout the FEC conditioning process. In brief, FEpic applies a Gaussian boundary filter on the decision plot that relates pixel intensity and distance between them. This filtering procedure establishes a map of local maxima (strongest emitter centers) laterally distributed across the imaging screen. Fig.4 shows an exemplary calculation of the number of emitters at $E_h$=70 MV/m which matches the simulation case in Fig.3. The full set of image processing results for $E_h$ ranging from 15 to 95 MV/m can be found in Appendix A. Dead pixels produced by cumulative x-ray damage were averaged out and smoothened as to not cause false emitter detection in the automated image processing.

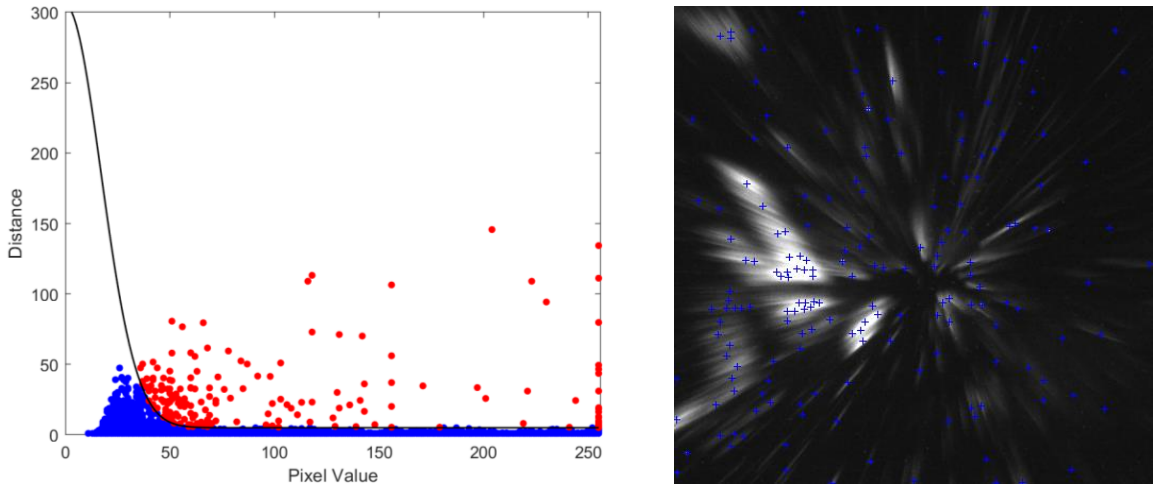

Figure 4: Image processing results for 70 MV/m: (a) decision plot with Gaussian boundary function, (b) processed image. All other decision plots and images with local maxima superimposed can be seen in Appendix A.

**Section V: Field Emission Cathode Performance**

Based on the high power conditioning methodology developed previously [8], the UNCD FEC was continuously conditioned from a turn-on field, determined to be 9 MV/m all the way up to 100 MV/m in increments of 5 MV/m. 9 MV/m was pinpointed to be the turn-on field as the electron emission signal was first detected on YAG1 at this field, though the charge on the Faraday cup was below the detection limit.

Q-E curves were taken as follows. The gradient was pushed up to the maximal intermediate target value termed $E_h$. The system was then kept at a given $E_h$ until the breakdown rate reduced to ~$10^{-4}$ per pulse. Every Q-E curve was then taken downwards with increments of 0.5 MV/m down to the point where no charge is detected by the Faraday cup. For every consecutive Q-E curve $E_h$ was 5 MV/m higher than $E_h$ for preceding Q-E curve.

The experimental session could not go above 70 MV/m as a breakdown event with a breakdown rate spiked instantaneously exceeding $10^{-1}$ per pulse occurred when the cathode was attempted to be run at 75 MV/m. Therefore, the first session was concluded, and a second session had to be carried out after the breakdown issue was identified. The *in situ* imaging system of the ACT identified the source of the intense breakdown sequence as a single location on the uncoated rounded molybdenum puck edge (see inset in Fig.5). The entire UNCD surface was determined to be unharmed by this major breakdown event. The FEC edge was refinished, and the FEC was reinserted into the gun for the second experimental session, demonstrating an additional depth of the UNCD FEC endurance capabilities.

During the second phase, the cathode was conditioned back up to 70 MV/m in a matter of two hours with only three breakdowns occurring during the entire process. Q-E curves were taken then from 70 MV/m to 95 MV/m in the same manner as was done in the first phase. 96±4 MV/m was the maximum field achievable due to power limited output of the klystron. While 95 MV/m was a targeted $E_h$, the actual mean gradient calculated from the measured input power was 96 MV/m with an error bar of 4 MV/m because the actual rf pulse envelope from the klystron is not an ideal rectangle. Fig.5 highlights that, despite the reinstallation process that involved micromachining and air exposure, the output charge remained at the same order of magnitude of 100 nC. Additionally, phase two experiments showed the steady increase of the output charge with increasing gradient. This is a record-breaking result for an rf injector where a field emission cathode produced 0.1 A of current at 100 MV/m.

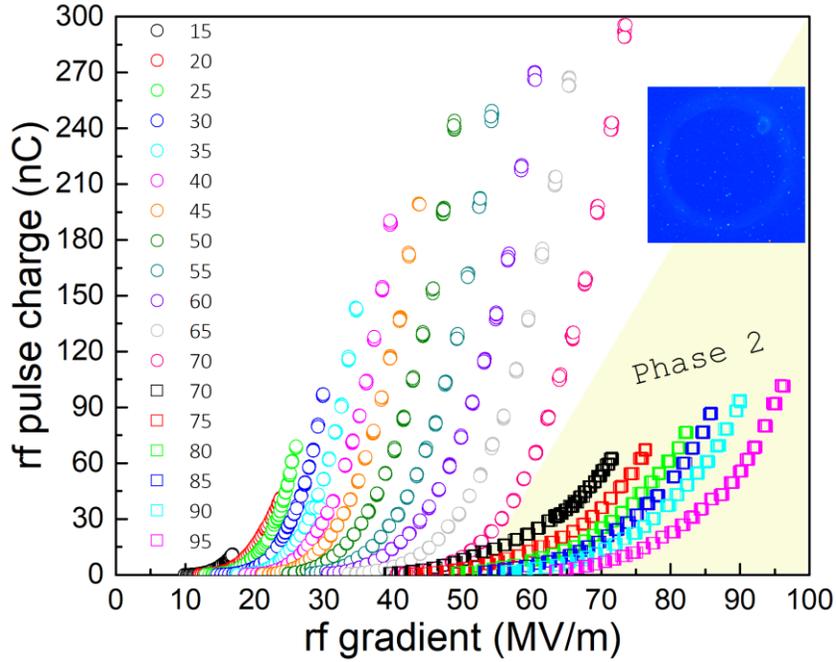

Figure 5: The Q-E curves for phase one and phase two separated into two regions corresponding to the experiment runs. In the legend, the numerical values label the maximal conditioning gradient achieved per Q-E curve (referred to as $E_h$). The inset shows *in situ* taken image of the major breakdown event, located on the outer edge of the cathode puck, that stopped the phase one session.

Raw data analysis and conversion to Q-E curves, as well as plotting and analysis in various coordinates, was accomplished using a data processing pipeline called FEbeam which is described in a great detail in Ref.[23]. FEbeam begins by taking the raw voltage signal waveforms for the diodes measuring forward power and reverse power and Faraday cup to calculate the Q-E curves which are then translated into Fowler-Nordheim coordinates by analyzing the rf envelope and finding its temporal length. The rf pulse length was 6 μs throughout the presented experiments. This resulted in a constant scaling factor when plotting either Q-E or I-E curves where I is the rf pulse current. As before, Fowler-Nordheim coordinates for rf case are $log_{10}\left(\frac{\bar{I}}{E^{2.5}}\right)$ versus $\left(\frac{1}{E}\right)$ [25] per modified FN relation

$$\overline{I_F(t)} = \frac{5.7 \times 10^{-12} \times 10^{4.52\phi^{-0.5}} A_e [\beta |E_c(t)|]^{2.5}}{\phi^{1.75}} \times \exp[-\frac{6.53 \times 10^9 \phi^{1.5}}{\beta |E_c(t)|}]$$

Unlike in previous rf cases [8,10], a significant divergence from classical Fowler-Nordheim law was revealed, which is obviously the result of the exceptionally high charge despite the duty cycle being extremely low, namely 6×10⁻⁶ at 1 Hz. The divergence is manifested by the presence of the knee point (previously discussed for dc case elsewhere [18]): two linear fits of a different slope exist intersecting at the knee point. Due to the large gradient incremental step (and thus, a smaller number of data points) a unique automated algorithm to retrieve the knee point was implemented in FEbeam. The algorithm determines the knee point location along with the relative ranges of the R² values of the iterative fitting for the line segments as depicted in Fig.6.

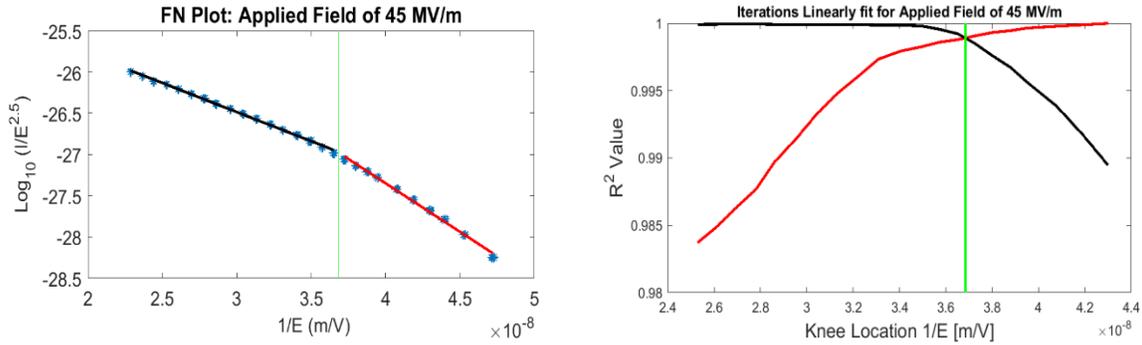

Figure 6: (a) Fowler Nordheim plot and (b) R² plot for 45 MV/m. The rest of the results can be seen in Appendix B.

After the knee point is selected, FEbeam performs FN fitting to two linear portions, separated by the knee point, in FN coordinates. Therefore, there are low gradient and high gradient portions of Q-E dependencies. For both portions, FEbeam then extracts the field enhancement factor (β), the local field on the cathode surface (β×$E_h$), and the effective emission area ($A_e$) all as a function of the maximum achieved conditioning field ($E_h$) per formalism below,

$$\begin{cases} \beta = \dfrac{-2.84 \times 10^9 \phi^{1.5}}{s} \\ A_e = \dfrac{10^{y_0} \phi^{1.75}}{5.7 \times 10^{-12} \times 10^{4.52\phi^{-0.5}} \beta} \end{cases},$$

where s and $y_o$ are the slope and the y-axis intercept of the linear dependence. The result summary is presented in Fig.7.

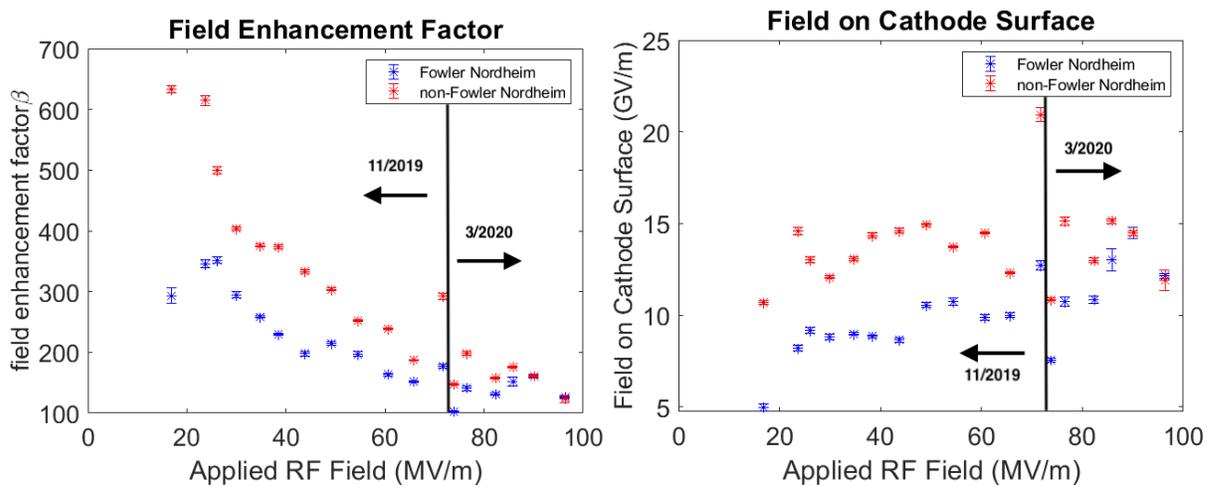

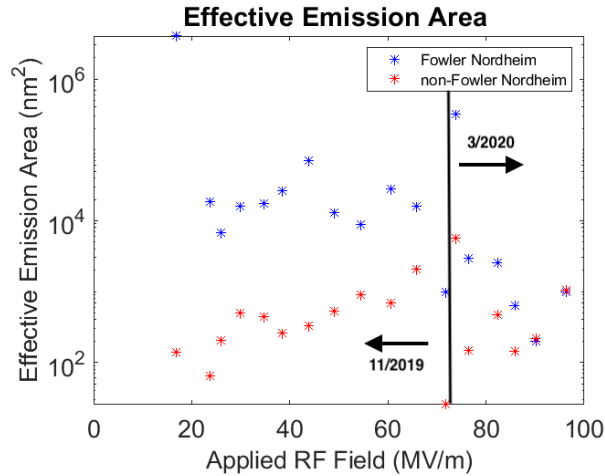

Figure 7: Field emission conditioning parameters: (a) field enhancement factor, (b) local field on the cathode surface, and (c) effective emission area for both the Fowler-Nordheim (low field) and non-Fowler-Nordheim (high field) regions.

High field data results in overestimating the local field, as it is known that a local field above 10 GV/m may be unphysical due to exceeding the limit set by the lattice force. This value includes diamond that breaks down at ~10 GV/m [26]. Experimentally, this would likely result in an immediate failure of the cathode due to breakdown induced runaway of the cathode material [27,28]. Since the cathode was not observed to behave this way, it is concluded that high field data overestimates the field enhancement factor. In contrast, low field data predicts the local field be below or at 10 GV/m. None of the datasets correctly predicted the emission area behavior. Even though incorrect, the effective emission area for the low field portion decreases with $E_h$ which is indicative of classical FN [18]. High field portion predicts the effective emission area to be a nearly constant value. Therefore, the low gradient portion is attributed to the classical metal-like FN behavior, while the high gradient portion is driven by a different physical mechanism which is discussed further.

### Section VI: Leading Hypothesis and Discussion

Generally, there are two competing hypotheses to explain deviation from FN behavior of planar FECs: 1) transit time and emission area limit charge transport and therefore emission (series resistor model) [18,29] and/or 2) space charge limited emission [29,30]. Our leading hypothesis is that it is space charge that affects the emission and causes the observed divergence. This is supported by a series of observations.

All measurements were done in fields stronger than $10^4$ V/cm. This means that, at all gradients, charge drift can be expected to be saturated [31]. The number of emitters quickly increased and remained near a constant value as obtained by *in situ* imaging at highest gradient points for every conditioning Q-E curve. Note that the effective emission area, as extracted from FN processing, was not used in space charge force calculation. It was the number of emitters extracted from imaging data processing that was used because it allowed for an observable independent of the QE curve fitting methodologies. For the purpose of this analysis, we were interested in the downstream imaging using YAG3 images as it enables the largest magnification and resolution (when coupled with the collimator at YAG2 position). The number of emitters

(or local maxima in terms of image processing terminology) on the cathode surface was determined by processing a set of 18 raw 16 bit images using FEpic [20] shown in Appendix A in its entirety.

It was assumed that each emitter calculated through image processing had the same area such that the total emission area is the number of emitters times the unit emitter area. The charge growth at every $E_h$ point slowed down but did not plateau out as it would be expected in the series resistor model. This confirms the leading role of the electric field outside (and not inside) the cathode. Fig.8 extends this representation by illustrating that the charge dynamics are driven by the electric field and not the emitters statistics (given the transport is saturated).

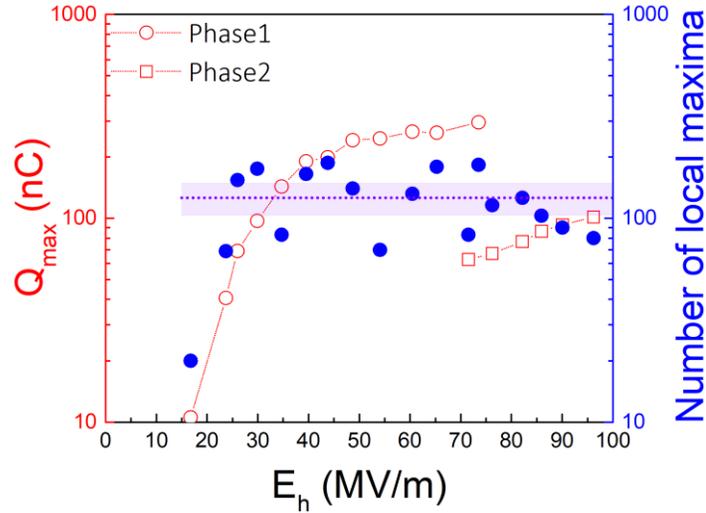

Figure 8: Maximum charge and the number of emitters determined at the maximum of every conditioning Q-E curve as functions of $E_h$. The purple horizontal area represents one standard deviation around the mean value of the emitter number count.

The knee point and the high field slope dynamically change. Fig.7a-c illustrates that high and low field results start converging for the gradients above 60 MV/m. Fig.9 maps the $R^2$ values from the iterative fitting knee point algorithm obtained from FN, $log_{10}\left(\frac{\bar{I}}{E^{2.5}}\right)$ versus $\left(\frac{1}{E}\right)$, and Millikan, $log_{10}(Q)$ versus $\left(\frac{1}{E}\right)$, representation of the experimental data. It indeed confirms that the knee point vanishes and both the high and low field regions converge to a single slope. Despite the charge increasing from 10 to 100 nC (Fig.8) as conditioning continued, the $R^2$ values quickly increased between 15 and 30 MV/m. In Phase 2, where the electric field was highest between 70 and 95 MV/m, $R^2$ remained at 0.999 implying a perfect linear relationship. These results indicate space charge domination at high charge and low gradient and essentially space charge suppression at gradients above 70 MV/m. This is indicated by the normalized space charge force $\propto \frac{Q}{N_e \cdot E_h^2}$ that is also plotted in Fig.9: the force is the inverse square of the field which quickly becomes the dominating term.

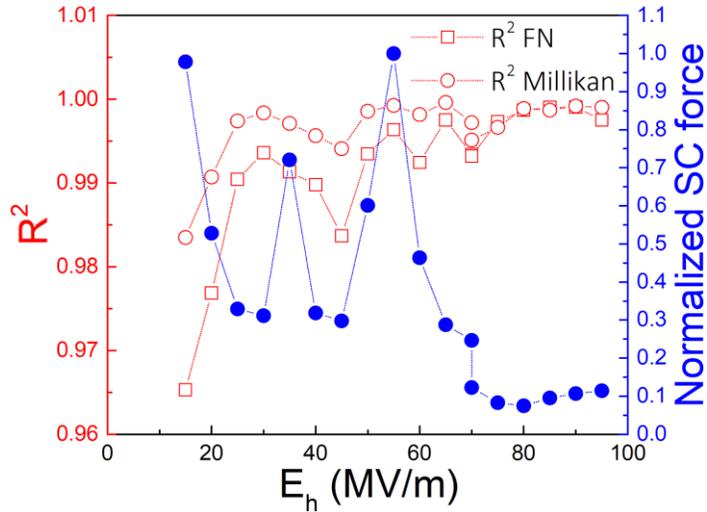

Figure 9: Combination of all trends for linearity of Fowler-Nordheim (open red square, left *y*-axis) and Millikan plots (red open circle, left *y*-axis) plotted against normalized space charge force (blue solid circles, right *y*-axis) which presents evidence for a space charge dominated Fowler-Nordheim regime.

Combining the field emission Q-E characteristics with the emitter statistics presents clear evidence that the divergence from the classical Fowler Nordheim regime is space charge dominated (but not limited) field emission. In space charge limited emission, the current or charge density would drop in FN coordinates. Having confirmed that the emitter count is roughly constant while the charge is a growing function of $E_h$ yields that the charge density still grows at a slower rate. Therefore, the new high field regime is termed space charge dominated FN emission (SCDFN). The difference between the space charge dominated and space charge limited emission is highlighted by the Millikan plot corresponding to 95 MV/m shown in Fig.10: the charge surpassed what the Child Langmuir limit predicts. A small parallel shift can be observed which is indicative of switching to space charge limited emission for high work function materials (>3.5 eV) [30]. The parallel shift starts forming at gradients $E_h$ above 70 MV/m, as can be found in Appendix B, and leads the field enhancements factors extracted from high low field portions of the Q-E curves to merge.

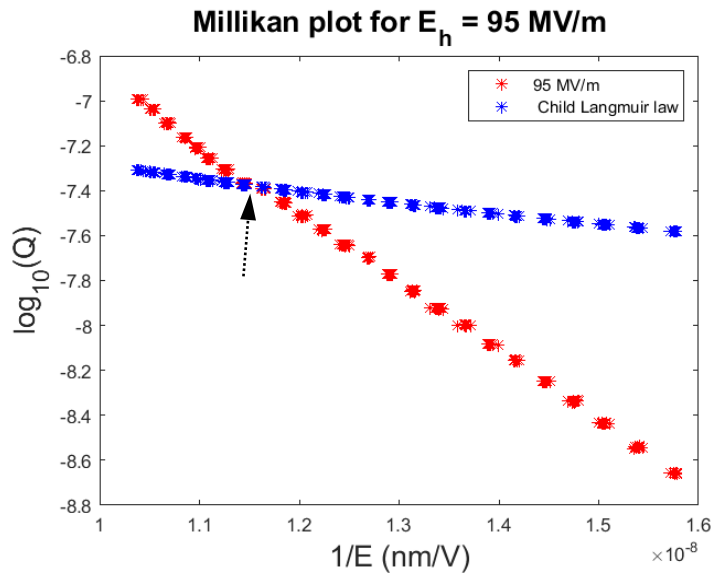

Figure 10: Millikan plot for 100 MV/m (red) showing the parallel shift denoted with an arrow and is indicative of a space charge dominated regime that differs from the Child Langmuir law (blue). Additional Millikan plots can be seen in Appendix B.

The reason behind surpassing the Child Langmuir limit merits further discussion. By its nature, UNCD has a periodic structure where sp2 grain boundaries alternate with sp3 diamond grains. Thus, added spatial inhomogeneity calls for description using a two-dimensional Child Langmuir problem to be considered. Luginsland *et al.* [32] demonstrated how the CL problem considered in 2D led to surpassing the classical 1D law of 2/3: 2D current was higher than 1D current by a factor of 4.

Finally, we estimate space charge emittance per our experimental conditions. First, we estimate the emission area to be ~$10^{-8}$ cm$^2$: such an area would bring the current density to ~$10^8$ A/cm$^2$ where the space charge effect starts to dominate. This yields an effective emitting radius of $10^{-4}$ cm: compare this value to the nominal FEC radius of 1 cm (much smaller value which is consistent with dc experiments [7,31]) and to the effective emitter radius $10^{-6}$–$10^{-5}$ cm obtained per FN formalism (one can note that area calculated from the high gradient portion provides quite a reasonable estimation). Taking into account that the effective emission window is 37 ps, defining $\sigma_z$, $\varepsilon_{sc}$ reads 0.033 mm×mrad and divergence angle reads 33 mrad for the rf pulse charge of 100 nC and surface gradient of 100 MV/m. For the given emission area radius, intrinsic emittance of UNCD can be estimated as $10^{-3}$ mm×mrad and a corresponding divergence angle of 1 mrad. Consequently, the brightness calculation provides an estimation of 6×$10^{14}$ A/m$^2$×rad$^2$. For most optimal rf designs, suppressing $\varepsilon_{rf}$ and enabling gradient at 300 MV/m level, field emission injectors could provide brightness close to $10^{16}$ A/m$^2$×rad$^2$, which would be a technological breakthrough.

**Section VII: Conclusion and Outlook**

Ultra-nano-crystalline field emission cathode operated at 100 MV/m in L-band in the space charge dominated regime was demonstrated. A charge of 100-300 nC per rf pulse (13 to 38 pC per rf cycle) was demonstrated. Output beam brightness near $10^{15}$ A/m$^2$×rad$^2$ is estimated for the given experimental conditions, and brightness near $10^{16}$ A/m$^2$×rad$^2$ is anticipated as a practical benchmark with further improvements of operating gradient of C- to X-band injector designs.

Further exploration and parametrization, in terms of the boundaries of charge and field, of the space charge dominated Fowler-Nordheim emission is underway. Immense gradient values enabled by microwave injectors allow for looking well beyond the 1D Child Langmuir physics, something extremely challenging with pulsed dc systems. Planar versus asperity geometries should provide insights into interplay between gradient and charge density effects in the space charge dominated/limited emission. In turn, understanding this new regime of field emission physics is crucial for informed improvement of the next generation of sources and injectors.

**Acknowledgments**

The work by Mitchell Schneider is supported by the U.S. Department of Energy Office of Science, High Energy Physics under Cooperative Agreement Award No. DE-SC0018362 and LANL through LANL LDRD 20200057DR under mentoring by John Lewellen and Evgenya I. Simakov. The work by Emily Jevarjian, Taha Posos and Sergey Baryshev was supported by the U.S. Department of Energy, Office of Science, Office of

High Energy Physics under Award No. DE-SC0020429. Tanvi Nikhar was funded by the College of Engineering at MSU under the Global Impact Initiative. The work at AWA was funded through the U.S. Department of Energy Office of Science under Contract No. DE-AC02-06CH11357. The authors would like to thank Manoel Conde and John Power for the continuing support of cathode research. We are grateful to Prof. Steve Lund (MSU) for useful discussions on the physics of space charge.

**References**

[1]     D. H. Dowell, *Topological cathodes: Controlling the space charge limit of electron emission using metamaterials*, Physical Review Accelerators and Beams **22** (2019).
[2]     K.-J. Kim, *Rf and space-charge effects in laser-driven rf electron guns*, Nuclear Instruments and Methods in Physics Research Section A: Accelerators, Spectrometers, Detectors and Associated Equipment **275**, 201 (1989).
[3]     L. Schächter, W. D. Kimura, and I. Ben-Zvi, *Ultrashort microbunch electron source*, AIP Conf. Proc. **1777**, 080013 (2016).
[4]     M. A. K. Othman, J. Picard, S. Schaub, V. A. Dolgashev, S. M. Lewis, J. Neilson, A. Haase, S. Jawla, B. Spataro, R. J. Temkin, S. Tantawi, and E. A. Nanni, *Experimental demonstration of externally driven millimeter-wave particle accelerator structure*, Applied Physics Letters **117**, 073502 (2020).
[5]     J. W. Lewellen and J. Noonan, *Field-emission cathode gating for rf electron guns*, Physical Review Special Topics - Accelerators and Beams **8** (2005).
[6]     X. Li, M. Li, L. Dan, Y. Liu, and C. Tang, *Cold cathode rf guns based study on field emission*, Physical Review Special Topics - Accelerators and Beams **16** (2013).
[7]     O. Chubenko, S. S. Baturin, K. K. Kovi, A. V. Sumant, and S. V. Baryshev, *Locally Resolved Electron Emission Area and Unified View of Field Emission from Ultrananocrystalline Diamond Films*, ACS Applied Materials & Interfaces **9**, 33229 (2017).
[8]     J. Shao, M. Schneider, G. Chen, T. Nikhar, K. K. Kovi, L. Spentzouris, E. Wisniewski, J. Power, M. Conde, W. Liu, and S. V. Baryshev, *High power conditioning and benchmarking of planar nitrogen-incorporated ultrananocrystalline diamond field emission electron source*, Physical Review Accelerators and Beams **22** (2019).
[9]     S. V. Baryshev, S. Antipov, J. Shao, C. Jing, K. J. Pérez Quintero, J. Qiu, W. Liu, W. Gai, A. D. Kanareykin, and A. V. Sumant, *Planar ultrananocrystalline diamond field emitter in accelerator radio frequency electron injector: Performance metrics*, Applied Physics Letters **105**, 203505 (2014).
[10]    S. V. Baryshev, E. Wang, C. Jing, V. Jabotinski, S. Antipov, A. D. Kanareykin, S. Belomestnykh, I. Ben-Zvi, L. Chen, Q. Wu, H. Li, and A. V. Sumant, *Cryogenic Operation of Planar Ultrananocrystalline Diamond Field Emission Source in SRF Injector*, Applied Physics Letters, 10.1063/5.0013172 (2021).
[11]    T. Nikhar, G. Adhikari, A. W. Schroeder, and S. V. Baryshev, *Evidence for Anti-Dowell-Schmerge Process in Photoemission from Diamond*, ArXiv **2011.00722** (2020).
[12]    H. Andrews, K. Nichols, D. Kim, E. I. Simakov, S. Antipov, G. Chen, M. Conde, D. Doran, G. Ha, W. Liu, J. Power, J. Shao, and E. Wisniewski, *Shaped Beams from Diamond Field-Emitter Array Cathodes*, IEEE Transactions on Plasma Science, 1 (2020).
[13]    K. E. Nichols, H. L. Andrews, D. Kim, E. I. Simakov, M. Conde, D. S. Doran, G. Ha, W. Liu, J. F. Power, J. Shao, C. Whiteford, E. E. Wisniewski, S. P. Antipov, and G. Chen, *Demonstration of transport of a patterned electron beam produced by diamond pyramid cathode in an rf gun*, Applied Physics Letters **116**, 023502 (2020).
[14]    R. L. Harniman, O. J. L. Fox, W. Janssen, S. Drijkoningen, K. Haenen, and P. W. May, *Direct observation of electron emission from grain boundaries in CVD diamond by PeakForce-controlled tunnelling atomic force microscopy*, Carbon **94**, 386 (2015).

[15]     J. B. Cui, J. Ristein, and L. Ley, *Low-threshold electron emission from diamond*, Physical Review B **60**, 16135 (1999).
[16]     T. Nikhar, R. Rechenberg, M. F. Becker, and S. V. Baryshev, *Dynamic graphitization of ultra-nano-crystalline diamond and its effects on material resistivity*, Journal of Applied Physics **128**, 235305 (2020).
[17]     J. Asmussen, T. A. Grotjohn, and T. Schuelke, in *Ultrananocrystalline Diamond* (Elsevier, 2012), pp. 53.
[18]     T. Y. Posos, S. B. Fairchild, J. Park, and S. V. Baryshev, *Field emission microscopy of carbon nanotube fibers: Evaluating and interpreting spatial emission*, Journal of Vacuum Science & Technology B **38**, 024006 (2020).
[19]     C. H. Ho, T. T. Yang, J. Y. Hwang, G. Y. Hsiung, S. Y. Ho, M. C. Lin, M. Conde, W. Gai, R. Konecny, J. Power, and P. Schoessow, *SRRC/ANL High Current L-band Single Cell Photocathode RF Gun*, EPAC Proceedings **TU33C**, 1441 (1998).
[20]     T. Y. Posos, O. Chubenko, and S. V. Baryshev, *Fast Pattern Recognition for Electron Emission Micrograph Analysis*, ArXiv **2012.03578** (2020).
[21]     E. Jevarjian, M. Schneider, and S. V. Baryshev, *FEgen (v.1): Field Emission Distribution Generator Freeware Based on Fowler-Nordheim Equation*, arXiv **2009.13046** (2020).
[22]     http://www.pulsar.nl/gpt/.
[23]     M. Schneider, E. Jevarjian, J. Shao, and S. V. Baryshev, *FEbeam: Cavity and Electron Emission Data Conversion, Processing and Analysis. A Freeware Toolkit for RF Injectors*, ArXiv **2012.10804** (2020).
[24]     G. Chen, G. Adhikari, L. Spentzouris, K. K. Kovi, S. Antipov, C. Jing, W. Andreas Schroeder, and S. V. Baryshev, *Mean transverse energy of ultrananocrystalline diamond photocathode*, Applied Physics Letters **114**, 093103 (2019).
[25]     J. W. Wang and G. A. Loew, *Field emission and rf breakdown in high-gradient room temperature linac structures*, SLAC-PUB-7684 (1997).
[26]     A. Hiraiwa and H. Kawarada, *Figure of merit of diamond power devices based on accurately estimated impact ionization processes*, Journal of Applied Physics **114**, 034506 (2013).
[27]     A. Kyritsakis, M. Veske, K. Eimre, V. Zadin, and F. Djurabekova, *Thermal runaway of metal nano-tips during intense electron emission*, Journal of Physics D: Applied Physics **51**, 225203 (2018).
[28]     S. S. Baturin, T. Nikhar, and S. V. Baryshev, *Field electron emission induced glow discharge in a nanodiamond vacuum diode*, Journal of Physics D: Applied Physics **52**, 325301 (2019).
[29]     J. W. Luginsland, A. Valfells, and Y. Y. Lau, *Effects of a series resistor on electron emission from a field emitter*, Applied Physics Letters **69**, 2770 (1996).
[30]     J. P. Barbour, W. W. Dolan, J. K. Trolan, E. E. Martin, and W. P. Dyke, *Space-Charge Effects in Field Emission*, Physical Review **92**, 45 (1953).
[31]     O. Chubenko, S. S. Baturin, and S. V. Baryshev, *Theoretical evaluation of electronic density-of-states and transport effects on field emission from n-type ultrananocrystalline diamond films*, Journal of Applied Physics **125**, 205303 (2019).
[32]     J. W. Luginsland, Y. Y. Lau, and R. M. Gilgenbach, *Two-Dimensional Child-Langmuir Law*, Physical Review Letters **77**, 4668 (1996).

## Appendix A

Shown here are decision plots and processed images indicating the number of local maxima found for each image.

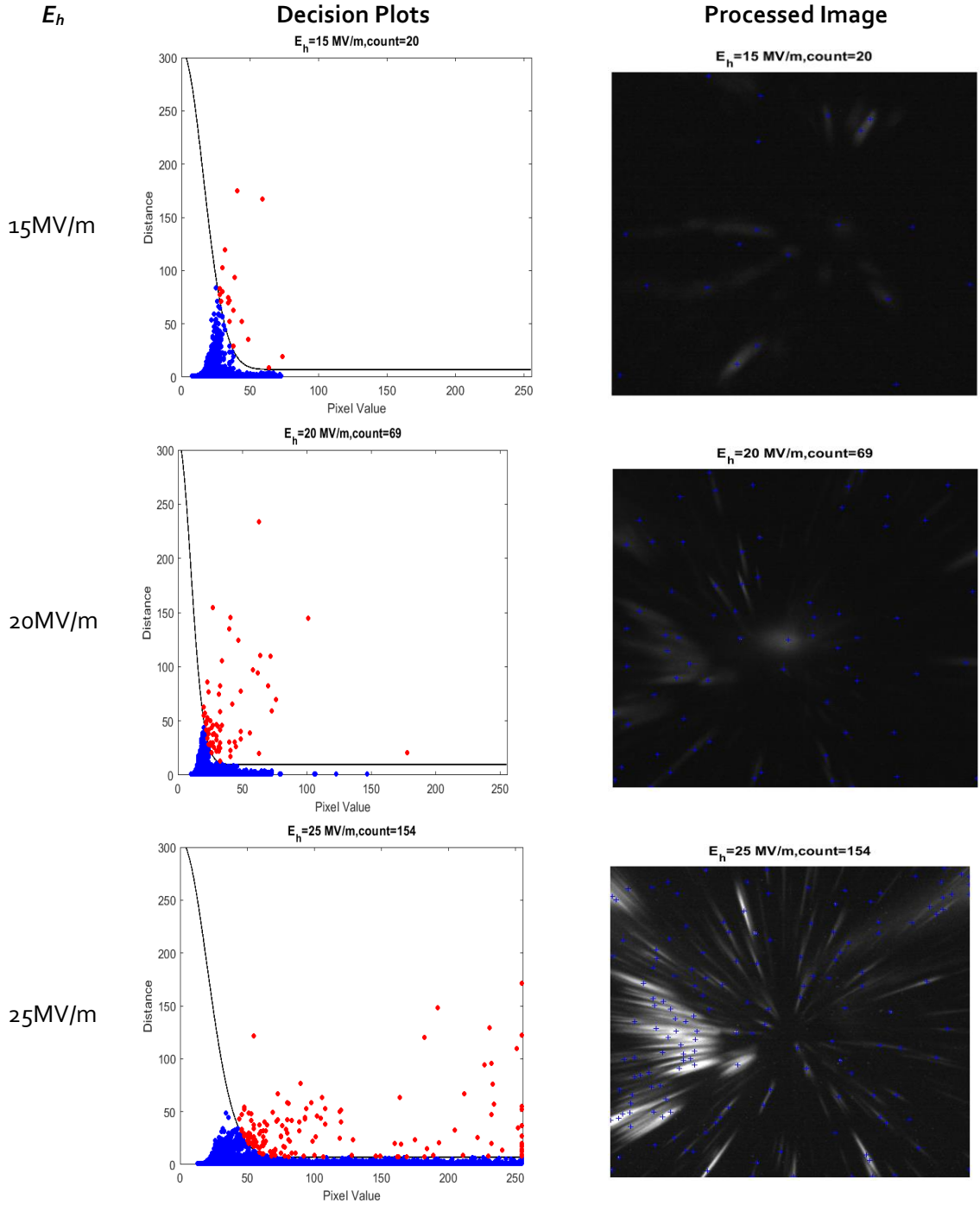

30MV/m 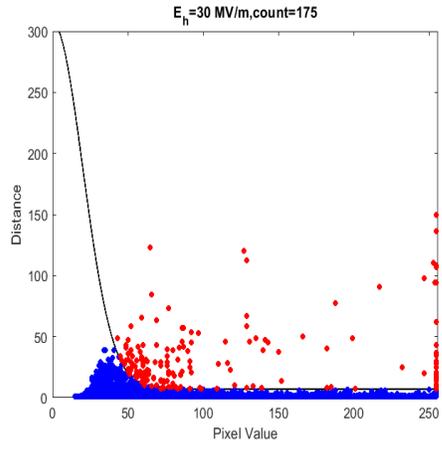 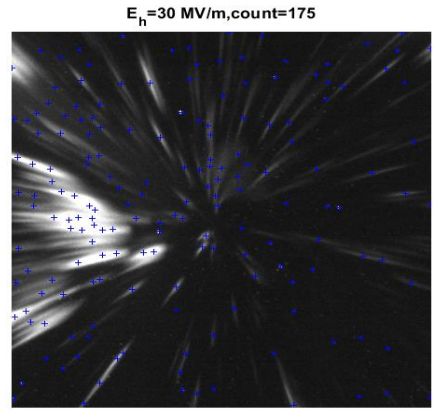

35MV/m 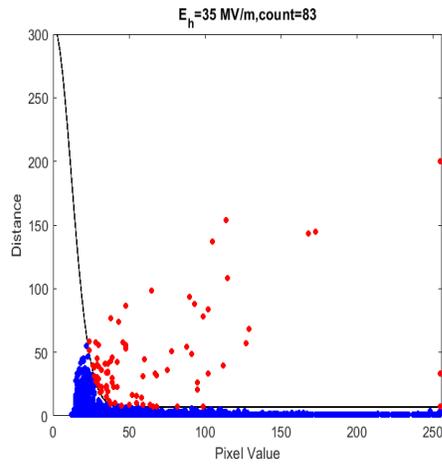 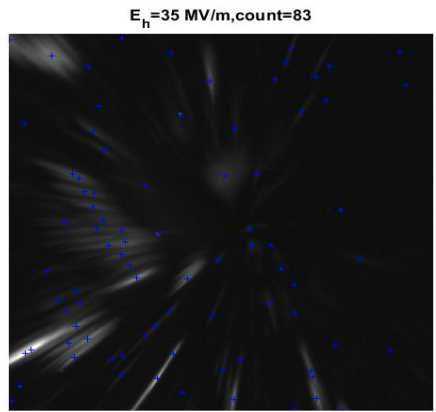

40MV/m 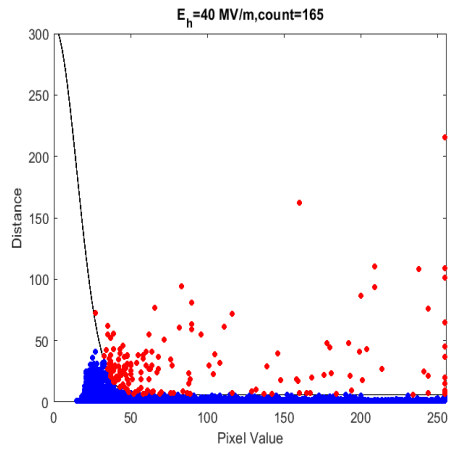 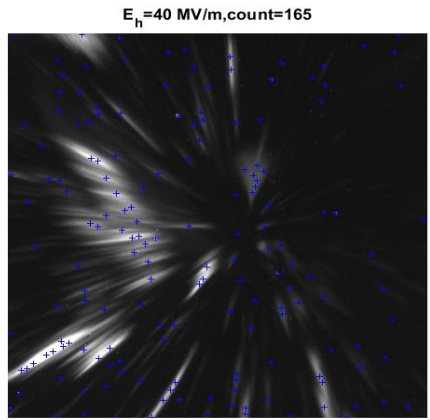

45MV/m 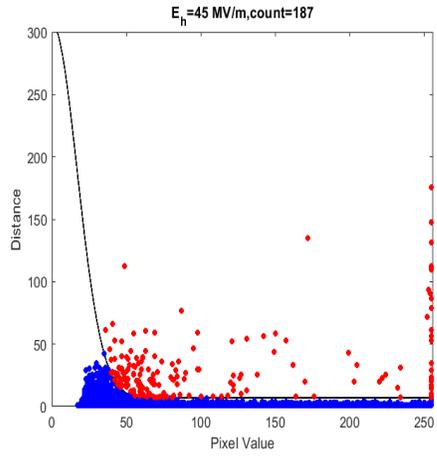 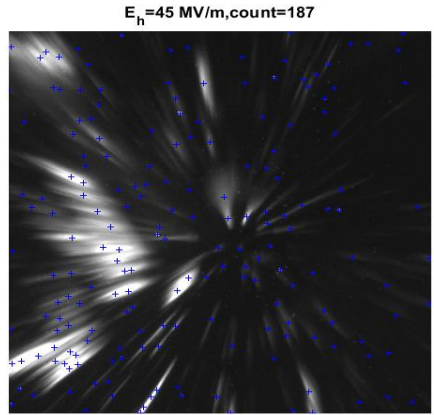

50MV/m 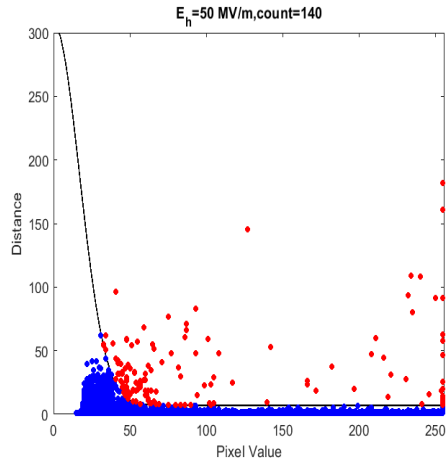 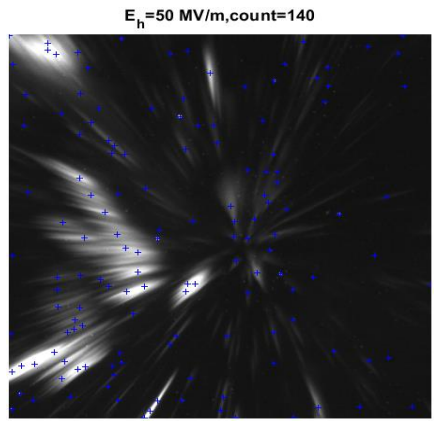

55MV/m 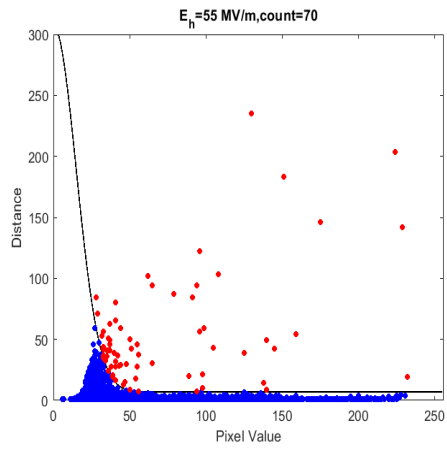 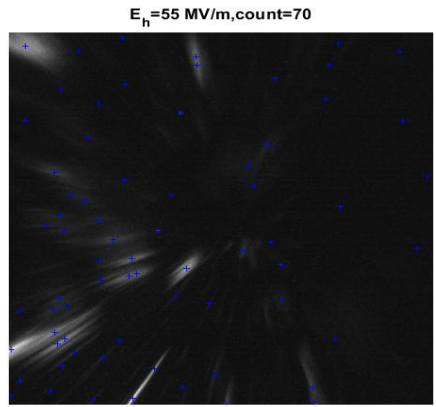

60MV/m
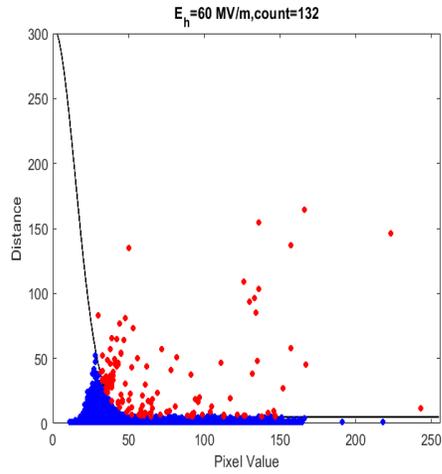 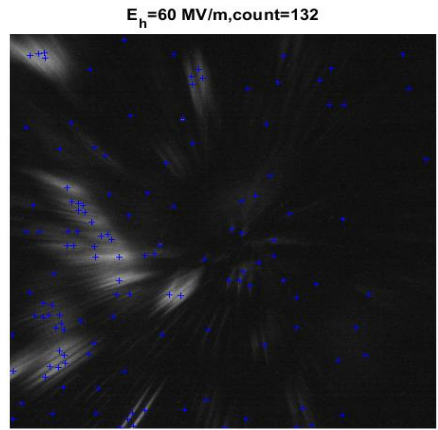

65MV/m
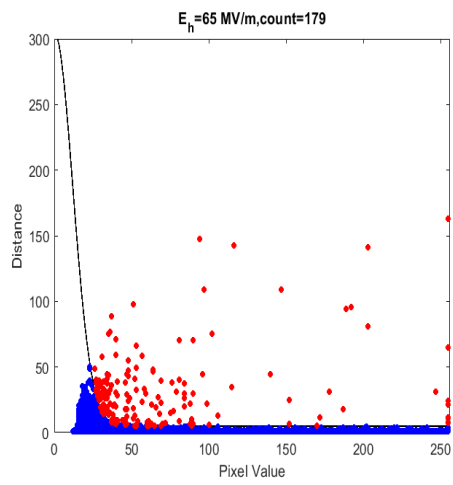 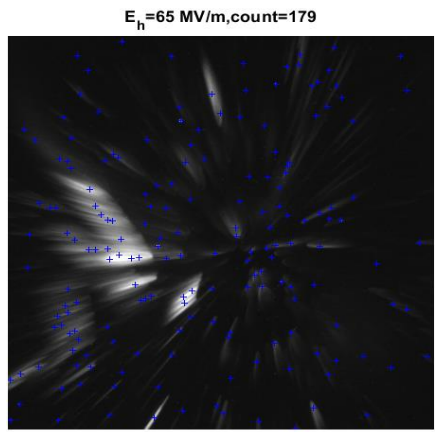

70MV/m Phase 1
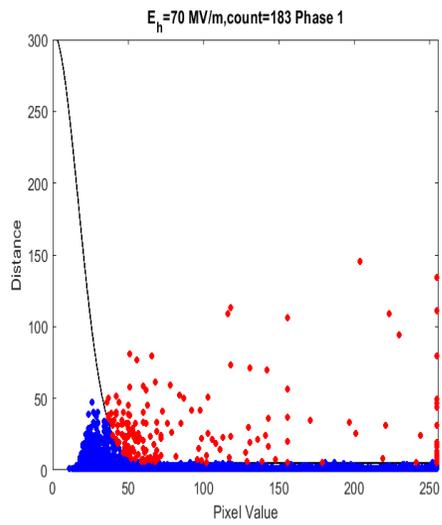 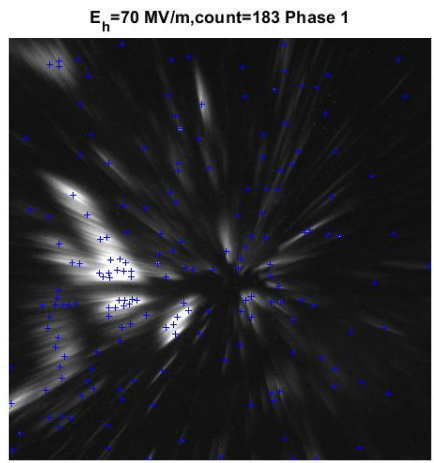

70MV/m
Phase 2

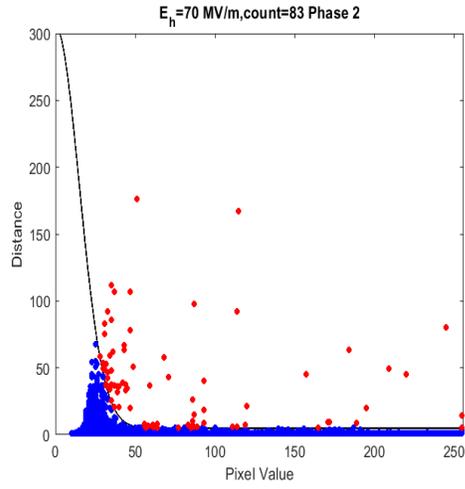
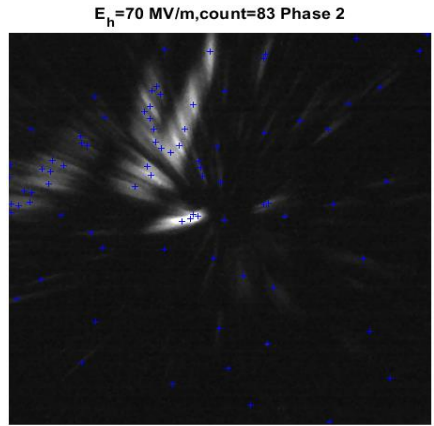

75MV/m

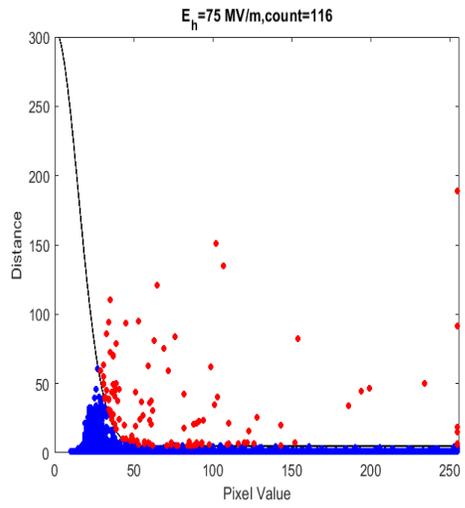
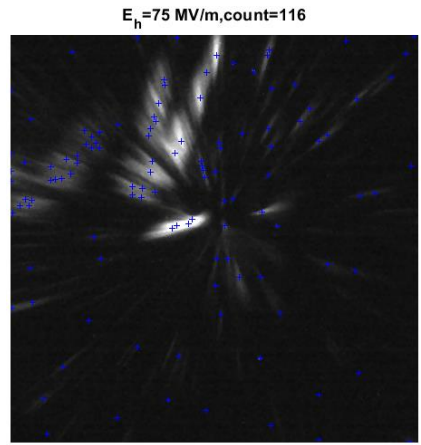

80MV/m

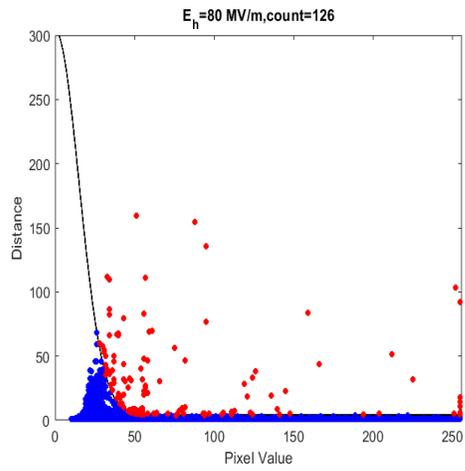
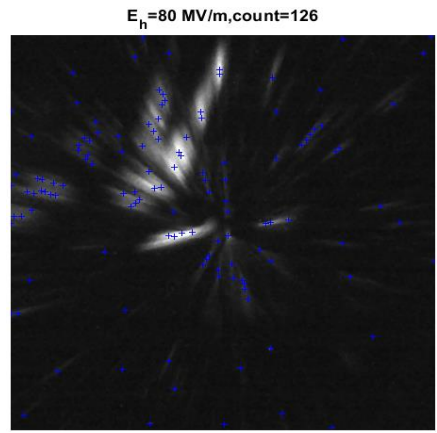

85MV/m
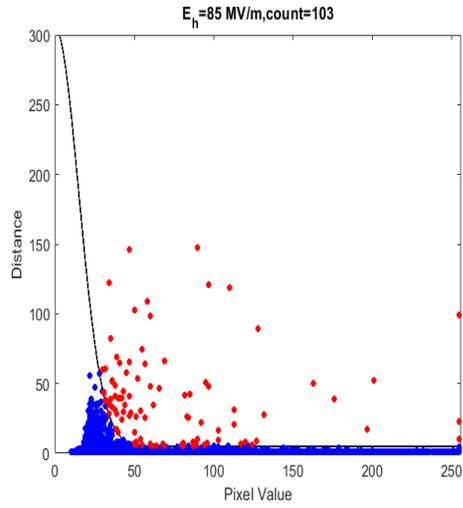 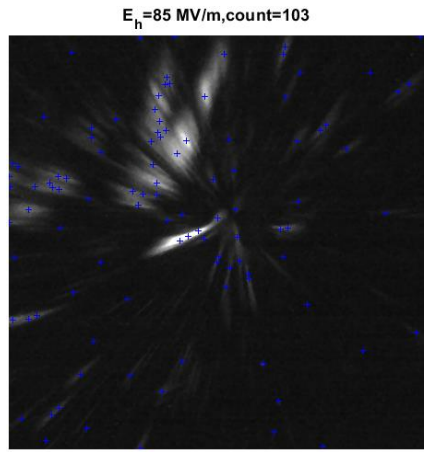

90MV/m
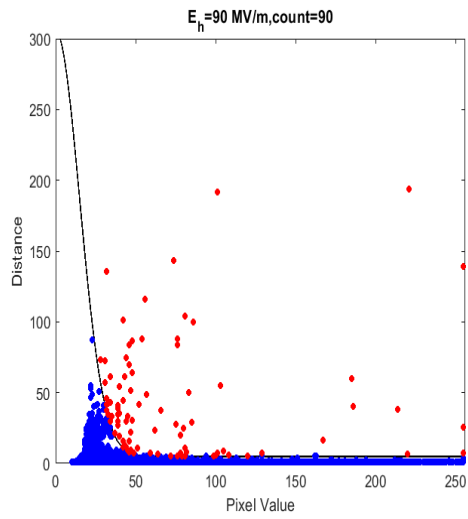 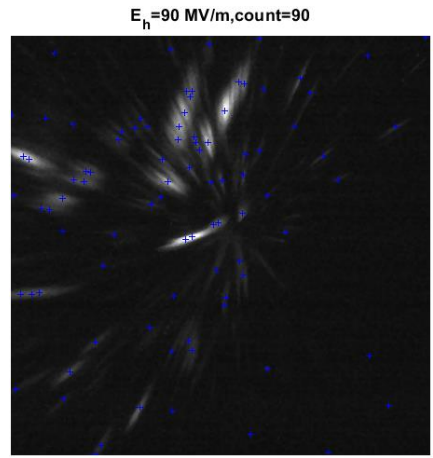

95MV/m
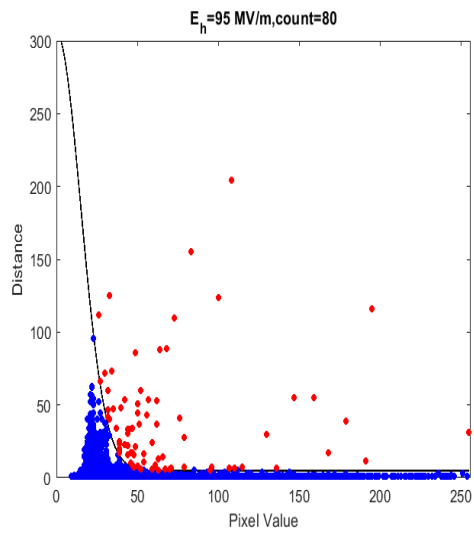 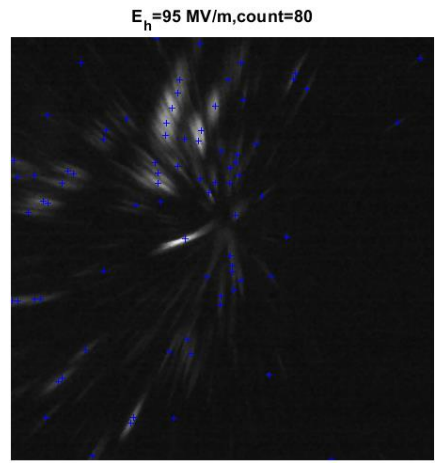

## Appendix B

Shown here are all the Fowler-Nordheim plots in I-E coordinates along with the corresponding R² plots used to find the knee point. Corresponding Millikan plots in Q-E coordinates are shown for comparison.

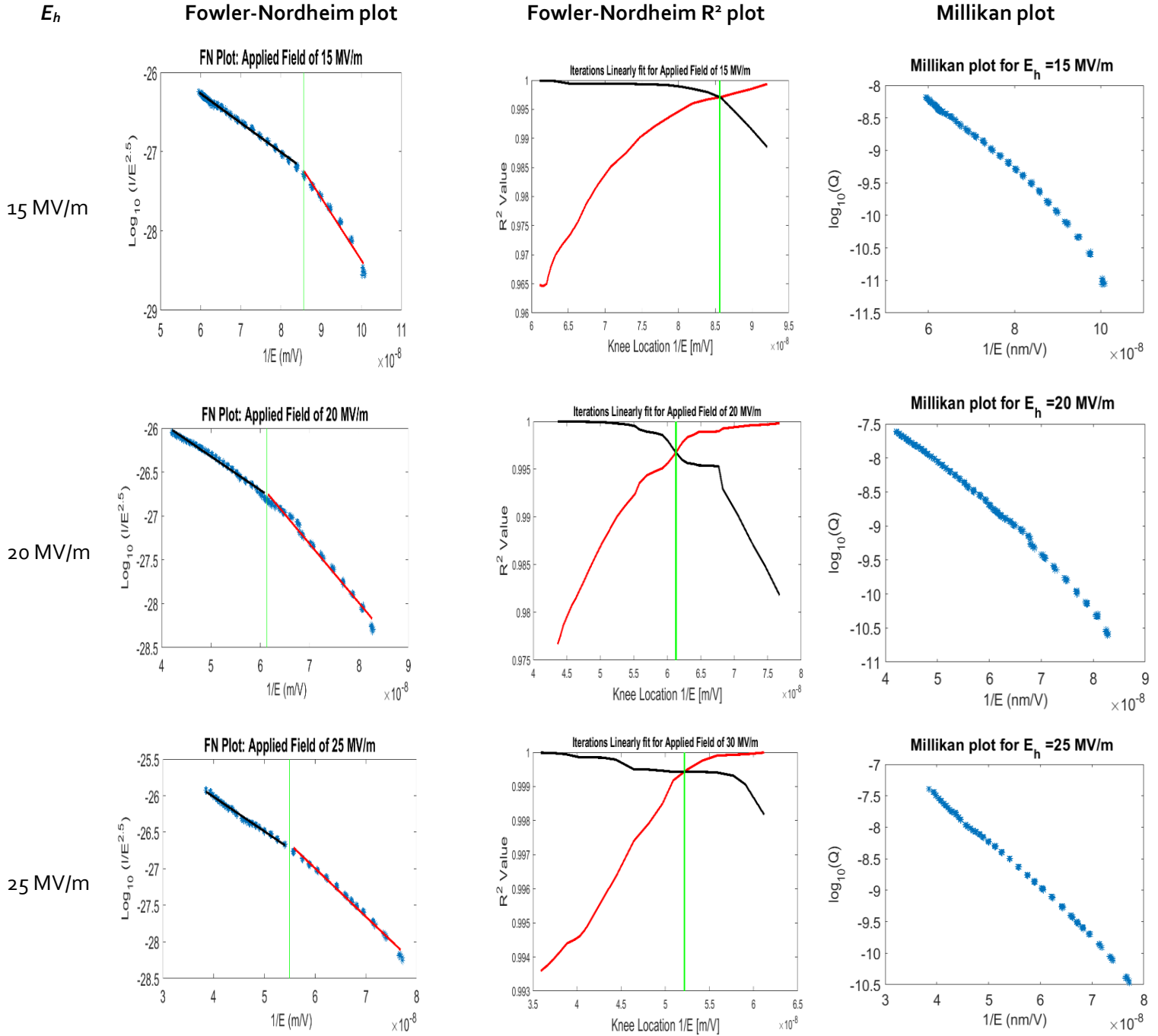

30 MV/m

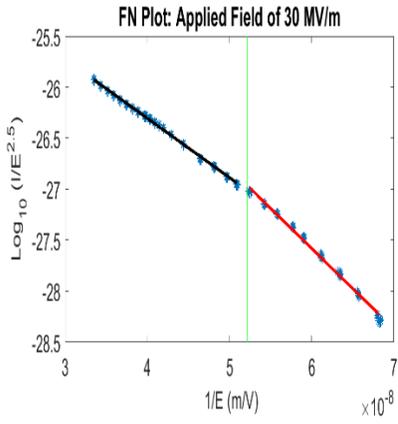 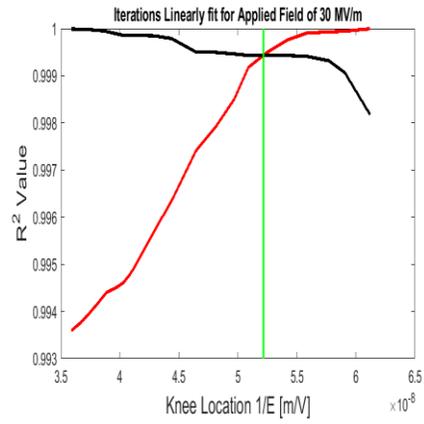 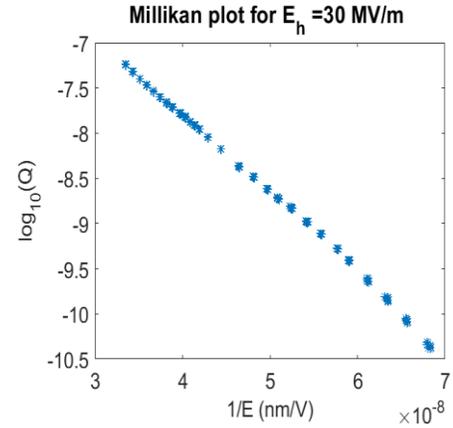

35 MV/m

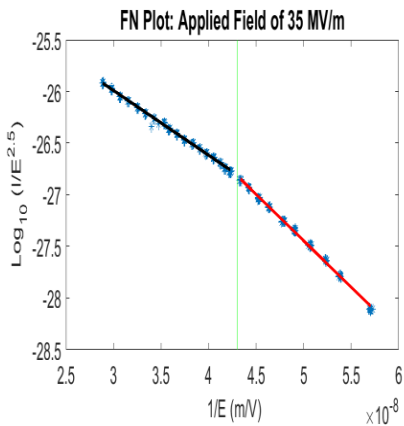 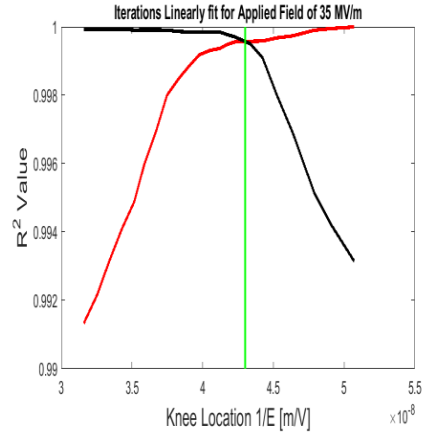 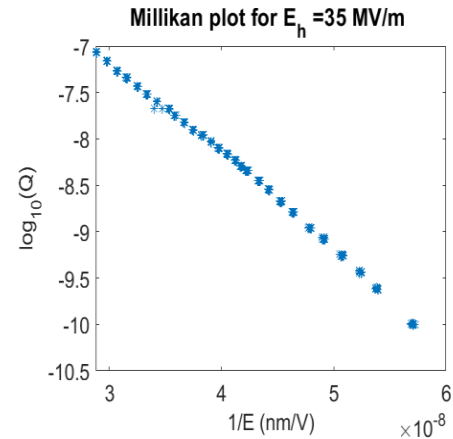

40 MV/m

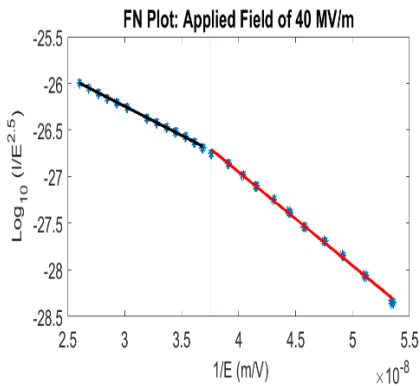 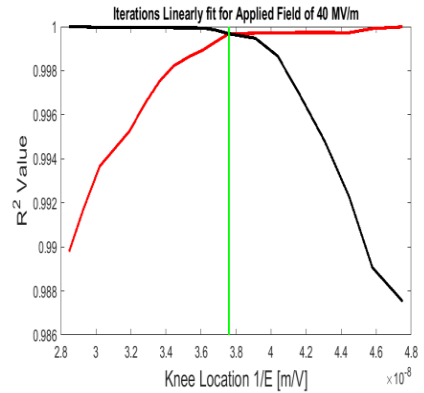 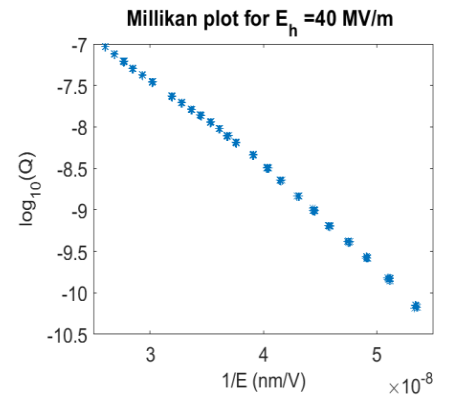

45 MV/m

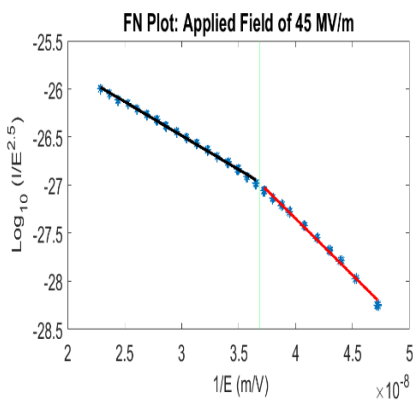 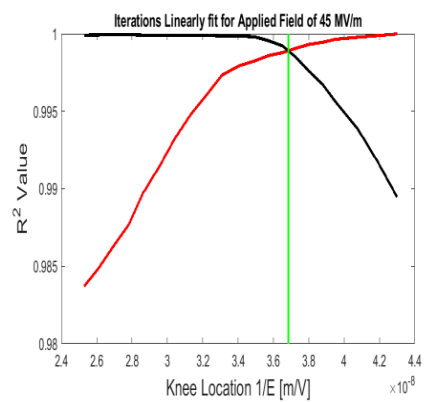 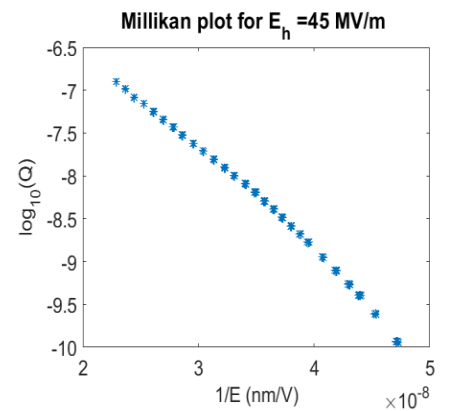

50 MV/m

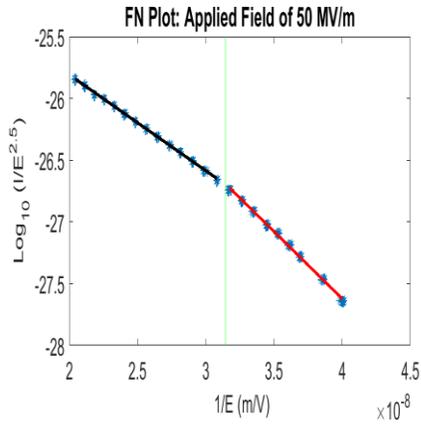 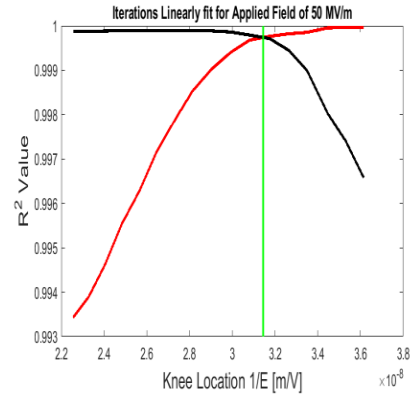 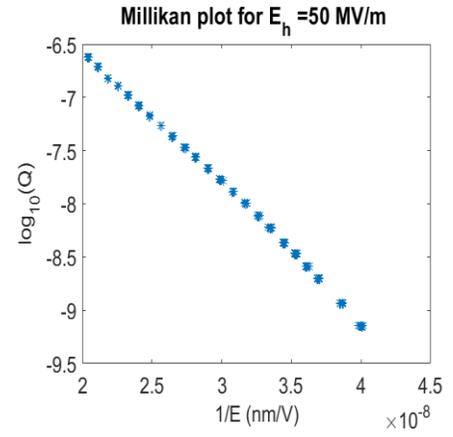

55 MV/m

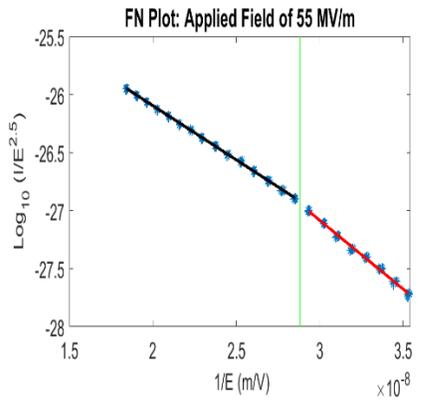 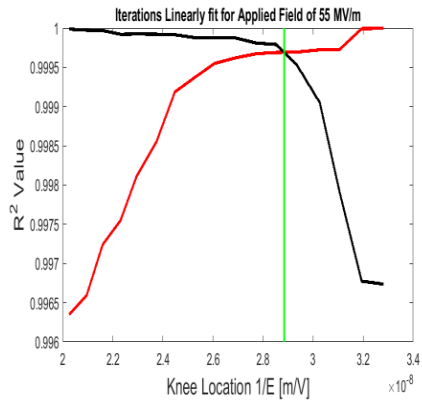 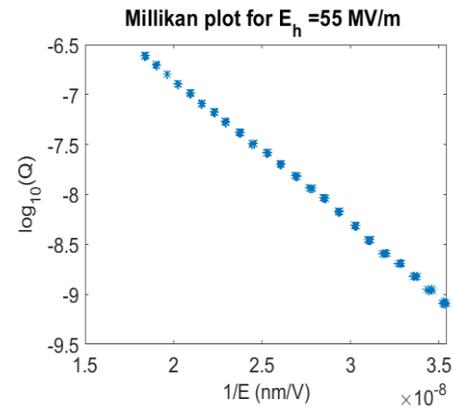

60 MV/m

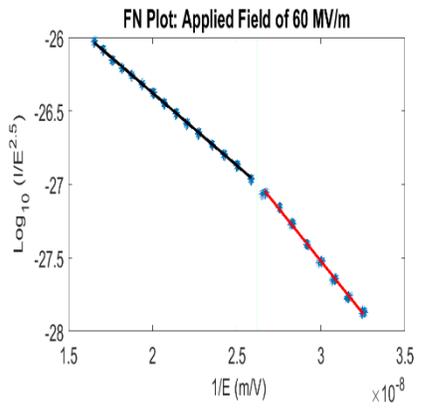 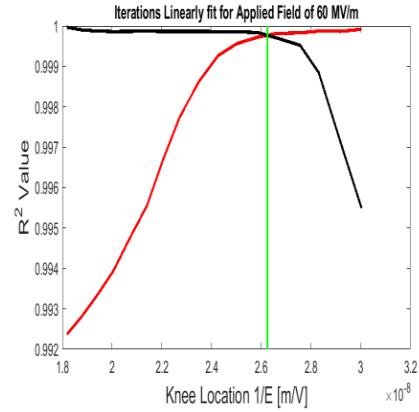 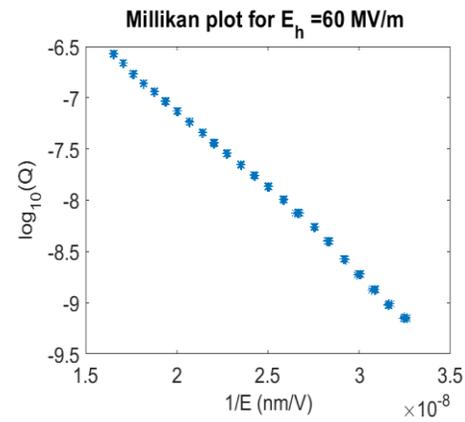

65 MV/m

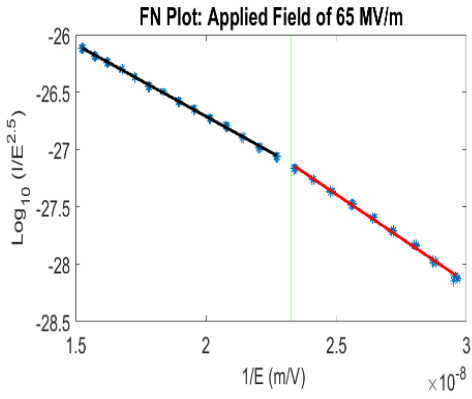 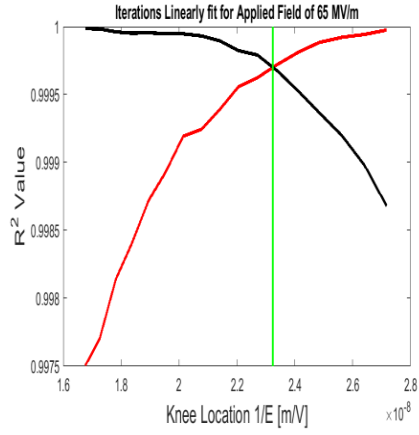 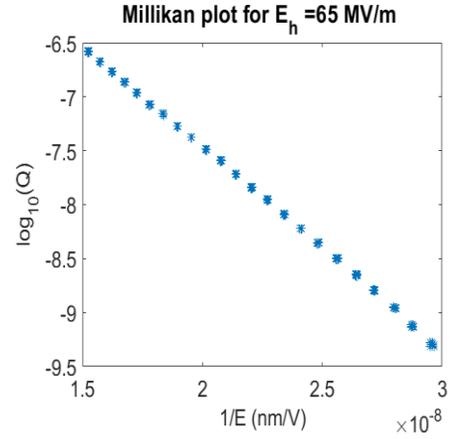

70 MV/m

Phase 1

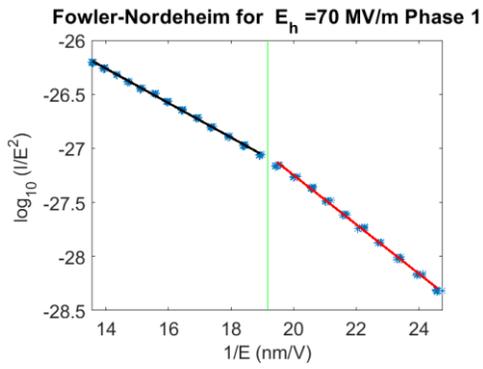 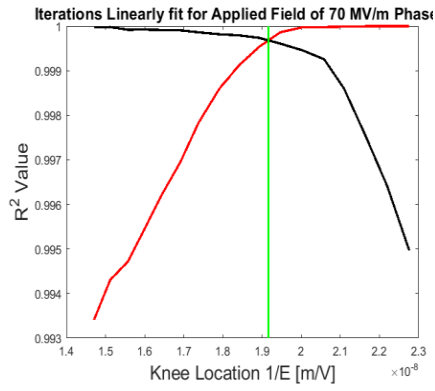 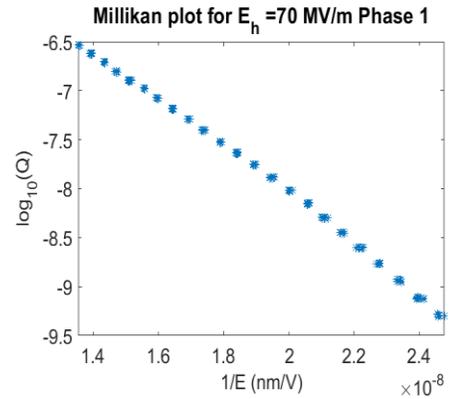

70 MV/m

Phase 2

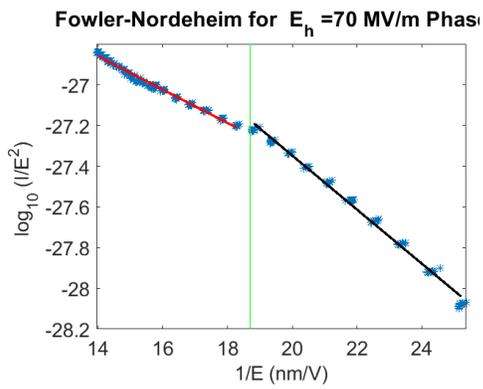 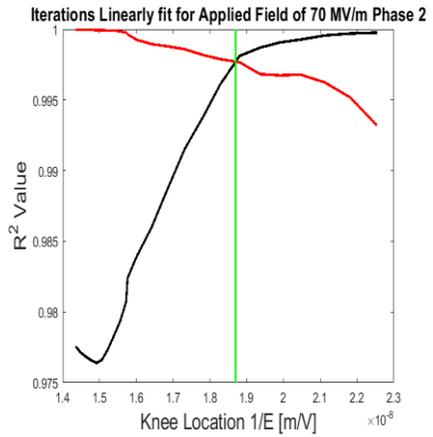 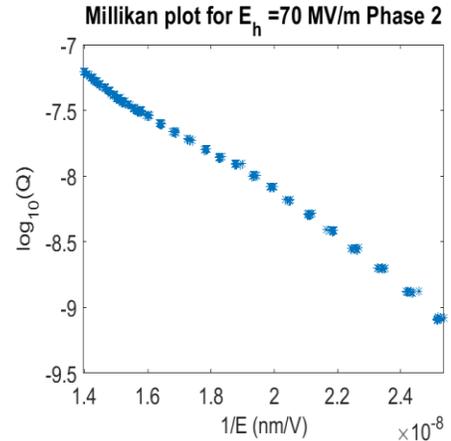

75 MV/m

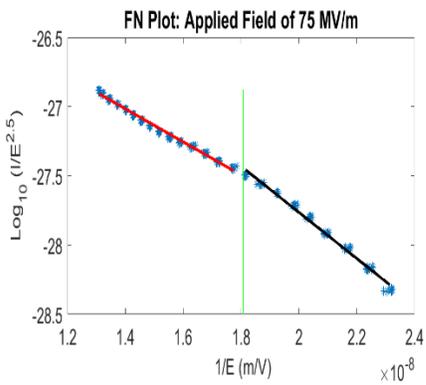 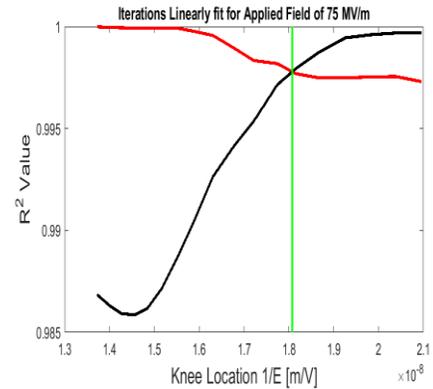 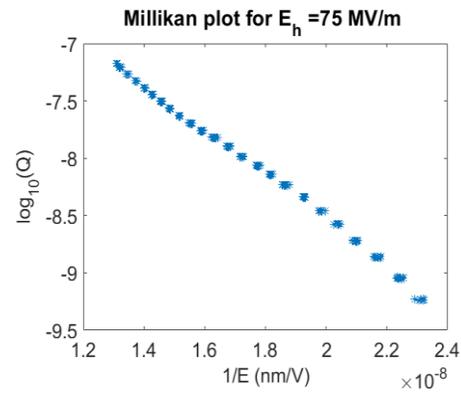

80 MV/m

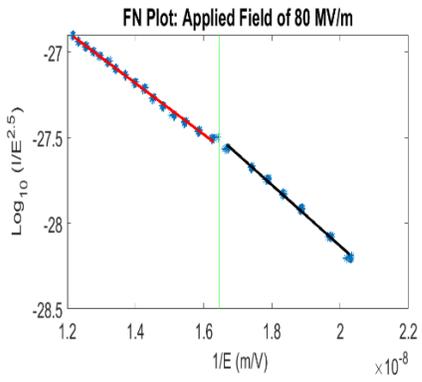 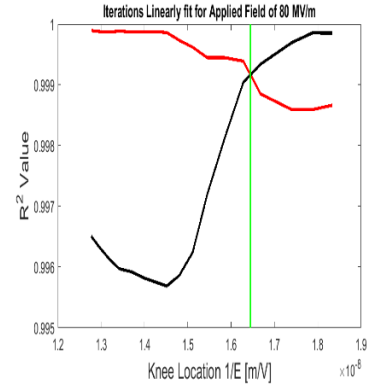 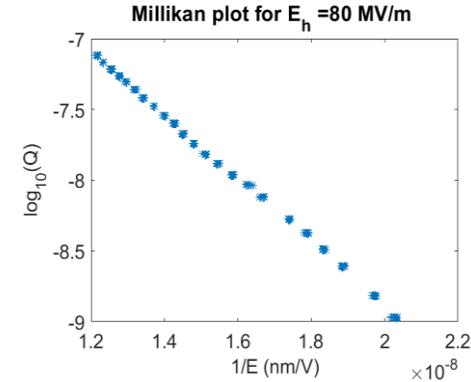

85 MV/m

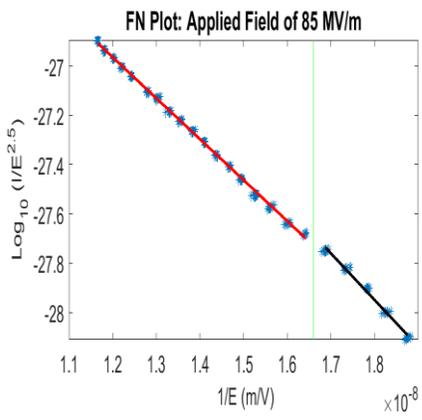 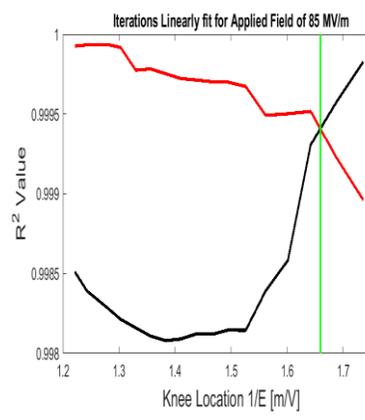 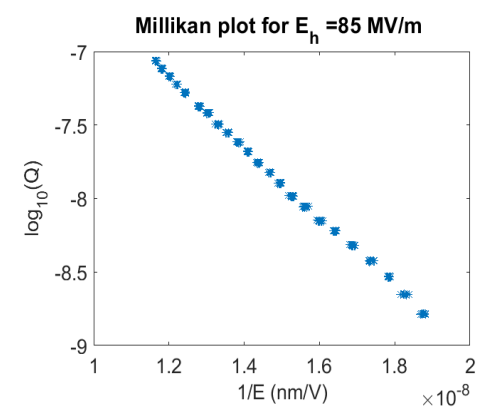

90 MV/m

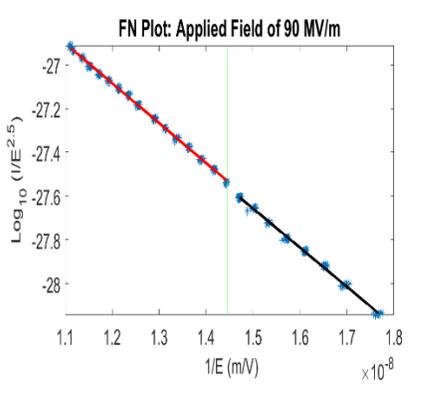 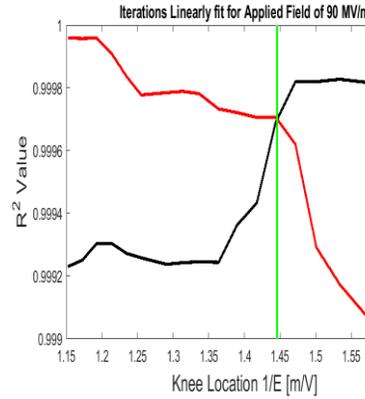 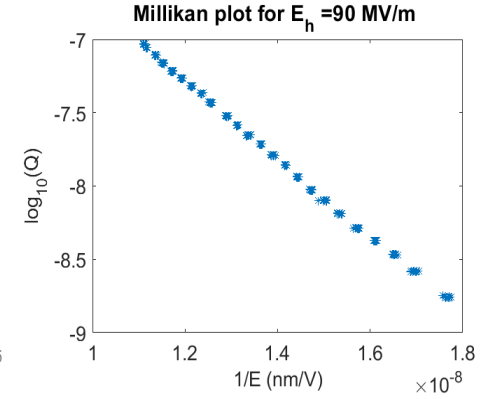

95 MV/m

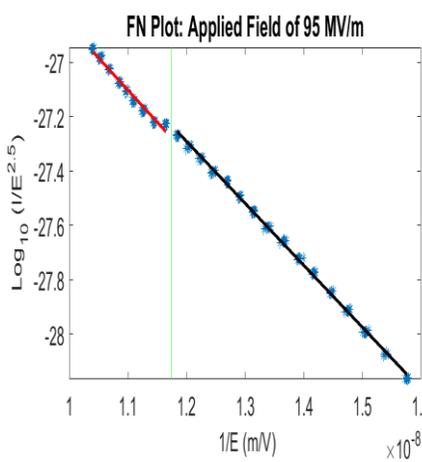 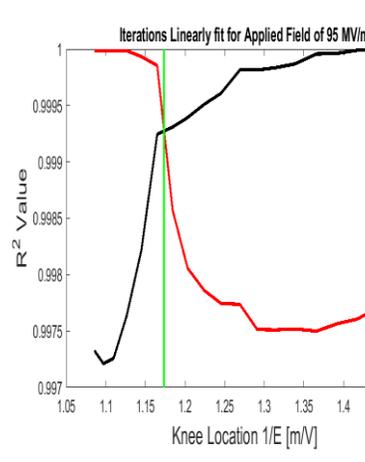 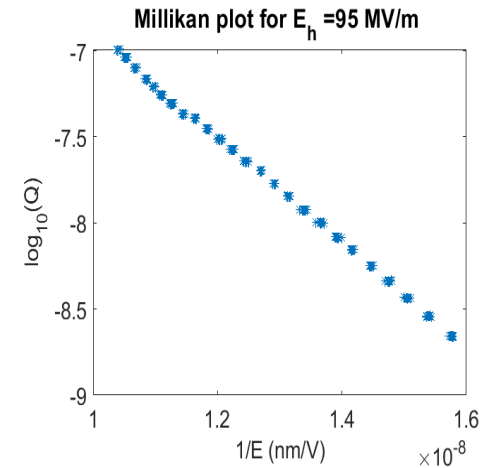